\documentclass{mystyle}

\begin{document}

\title{Theory of CP Violation in the B-Meson System}

\author{Matthias Neubert}

\address{Newman Laboratory of Nuclear Studies, Cornell University, 
Ithaca, NY 14853, USA\\
E-mail: neubert@mail.lns.cornell.edu}

\twocolumn[\maketitle\abstract{
Recent developments in the theory of CP violation in the $B$-meson 
system are reviewed, with focus on the determination of $\sin2\beta$ 
from $B\to J/\psi\,K$ decays, its implications for tests of the Standard 
Model and searches for New Physics, and the determination of $\gamma$ 
and $\alpha$ from charmless hadronic $B$ decays.}]

\section{Introduction}

The phenomenon of CP violation is one of the most intriguing aspects of
modern physics, with far-reaching implications for the microscopic 
world as well as for the macrocosmos. CP violation means that there is 
a fundamental difference between the interactions of matter and 
anti-matter, which in conjunction with the CPT theorem implies a 
microscopic violation of time-reversal invariance. CP violation is also 
responsible for the observed asymmetry in the abundance of matter and 
anti-matter in the Universe, which is a prerequisite for our existence.

The Standard Model (SM) of particle physics provides us with a 
parameterization of CP violation but does not explain its origin in a 
satisfactory way. In fact, CP violation may occur in three sectors of the
SM: in the quark sector via the phase of the Cabibbo--Kobayashi--Maskawa 
(CKM) matrix, in the lepton sector via the phases of the neutrino mixing 
matrix, and in the strong interactions via the parameter 
$\theta_{\rm QCD}$. CP violation in the quark sector has been studied in 
some detail and is the subject of this talk. The nonobservation of CP 
violation in the strong interactions is a mystery (the ``strong CP 
puzzle''\cite{Quinn:2001vt}), whose explanation requires physics beyond 
the SM (such as a Peccei--Quinn symmetry, axions, etc.). CP violation 
in the neutrino sector has not yet been explored experimentally.

The discovery of CP violation in the $B$ system, as reported this summer
by the BaBar\cite{Aubert:2001nu} and Belle\cite{Abe:2001xe} 
Collaborations, is a triumph for the Standard Model. There is now 
compelling evidence that the phase of the CKM matrix correctly explains 
the pattern of CP-violating effects in mixing and weak decays of kaons, 
charm and beauty hadrons. Specifically, the CKM mechanism explains why 
CP violation is a small effect in $K$--$\bar K$ mixing ($\epsilon_K$) 
and $K\to\pi\pi$ decays ($\epsilon'/\epsilon$), why CP-violating effects 
in charm physics are below the sensitivity of present experiments, and 
why CP violation is small in $B$--$\bar B$ mixing ($\epsilon_B$) but 
large in the interference of mixing and decay in $B\to J/\psi\,K$ 
($\sin2\beta_{\psi K}$). The significance of the $\sin2\beta_{\psi K}$ 
measurement is that for the first time a large CP asymmetry has been 
observed, proving that CP is not an approximate symmetry of Nature. 
Rather, the smallness of CP violation outside the $B$ system simply 
reflects the hierarchy of CKM matrix elements. 

\begin{figure}
\centerline{\includegraphics[width=0.48\textwidth]{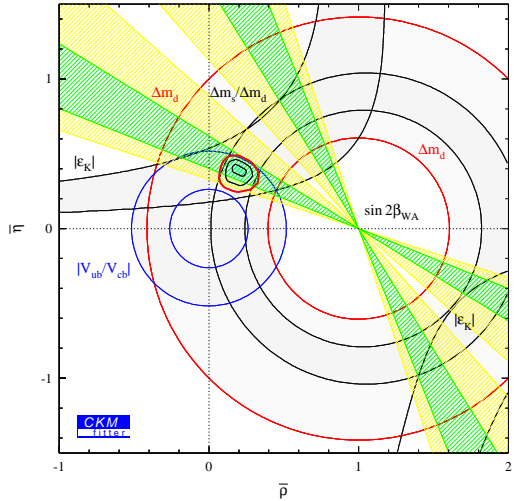}}
\caption{Summary of standard constraints and global fit in the 
$(\bar\rho,\bar\eta)$ plane.\protect\cite{Hocker:2001xe}}
\label{fig:CKMfit}
\end{figure}

Besides CP violation, the CKM mechanism explains a vast variety of 
flavor-changing processes, including leptonic decays, semileptonic 
decays (from which the magnitudes of most CKM elements are determined), 
nonleptonic decays, rare loop processes, and mixing amplitudes. The CKM
matrix is a unitary matrix in flavor space, which relates the mass 
eigenstates of the down-type quarks with the interaction eigenstates 
that are involved in flavor-changing weak transitions. It has a 
hierarchical structure, and (with the standard phase conventions) the 
CP-violating phase appears in the smallest matrix elements, $V_{ub}$ and 
$V_{td}$. The CKM matrix is described by four observable parameters, 
which can be taken to be the parameters of the Wolfenstein 
parameterization. Two of them, $\lambda=0.222\pm 0.004$ and 
$A=0.83\pm 0.07$ (at 95\% confidence level) are rather accurately known, 
whereas the other two, $\bar\rho$ and $i\bar\eta$, are more uncertain. 
A convenient way of summarizing the existing information on $\bar\rho$ 
and $\bar\eta$ is to represent the unitarity relation 
$V_{ub}^* V_{ud}+V_{cb}^* V_{cd}+V_{tb}^* V_{td}=0$ as a triangle in 
the complex plane. If the triangle is rescaled such that is has base of 
unit length, then the coordinates of the apex are given by 
$(\bar\rho,\bar\eta)$. The angles of the unitarity triangle are related 
to CP violation. Figure~\ref{fig:CKMfit} shows an 
example\cite{Hocker:2001xe} of a recent global analysis of the unitarity 
triangle, combining measurements of $|V_{cb}|$ and $|V_{ub}|$ in 
semileptonic $B$ decays, $|V_{td}|$ in $B$--$\bar B$ mixing, and the 
CP-violating phase of $V_{td}^2$ in $K$--$\bar K$ mixing and 
$B\to J/\psi\,K$ decays. The values obtained at 95\% confidence level 
are $\bar\rho=0.21\pm 0.12$ and $\bar\eta=0.38\pm 0.11$. The 
corresponding results for the angles of the unitarity triangle are 
$\sin 2\beta=0.74\pm 0.14$, $\sin2\alpha=-0.14\pm 0.57$, and 
$\gamma=(62\pm 15)^\circ$. These studies have established the existence 
of a CP-violating phase in the top sector of the CKM matrix, i.e., the
fact that $\mbox{Im}(V_{td}^2)\propto\bar\eta\ne 0$.

\boldmath
\section{Determination of $\sin2\beta_{\psi K}$}
\unboldmath

In decays of neutral $B$ mesons into a CP eigenstate $f_{\rm CP}$, an 
observable CP asymmetry can arise from the interference of the 
amplitudes for decays with an without $B$--$\bar B$ mixing, i.e., from 
the fact that the amplitudes for $B^0\to f_{\rm CP}$ and  
$B^0\to\bar B^0\to f_{\rm CP}$ must be added coherently. The resulting
time-dependent asymmetry is given by 
\[
   A_{\rm CP}(t) = 
   \frac{\Gamma(\bar B^0(t)\!\to\! f_{\rm CP})
         -\Gamma(B^0(t)\!\to\! f_{\rm CP})}
        {\Gamma(\bar B^0(t)\!\to\! f_{\rm CP})
         +\Gamma(B^0(t)\!\to\! f_{\rm CP})} 
\]
\vspace{-0.5truecm}
\[
   = \frac{2\mbox{Im}(\hat\lambda)}{1+|\hat\lambda|^2}
   \sin(\Delta m_d t)
   - \frac{1-|\hat\lambda|^2}{1+|\hat\lambda|^2} \cos(\Delta m_d t) \,,
\]
where $\hat\lambda=e^{i\phi_d} \bar A/A$, $\phi_d$ is the $B$--$\bar B$ 
mixing phase (which in the SM equals $-2\beta$), and $A\,(\bar A)$ 
denotes the $B^0\,(\bar B^0)\to f_{\rm CP}$ decay amplitude. If the 
amplitude is dominated by a single weak phase $\phi_A$, then 
$|\hat\lambda|\simeq 1$ and
\[
   A_{\rm CP}(t) \simeq \eta_{f_{\rm CP}} \sin(\phi_d-2\phi_A) 
   \sin(\Delta m_d t) \,,
\]
where $\eta_{f_{\rm CP}}=\pm 1$ is the CP signature of the final state.

\begin{figure}
\centerline{\includegraphics[width=0.48\textwidth]{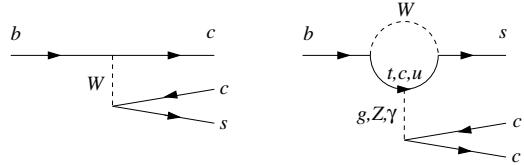}}
\caption{Tree and penguin topologies in $B\to J/\psi\,K$ decays.}
\label{fig:topol}
\end{figure}

The ``golden mode'' $B\to J/\psi\,K_S$ (for which $\eta_{\psi K_S}=-1$)
is based on $b\to c\bar c s$ transitions, which in the SM can proceed 
via tree or penguin topologies, as shown in Figure~\ref{fig:topol}. To 
an excellent approximation the decay amplitude for this process is real. 
A weak phase is introduced only through components of the up- and 
top-quark penguin diagrams that are strongly CKM suppressed. 
Parametrically, the ``penguin pollution'' to the weak phase from these 
effects is of order $\Delta\phi_A\sim\lambda^2(P/T)\sim 1\%$, where 
$P/T\sim 0.2$ is the tree-to-penguin ratio. It follows that
$A_{\rm CP}(t)\simeq\sin2\beta\sin(\Delta m_d t)$ with an accuracy of 
about 1\%. 

This summer, the BaBar and Belle Collaborations have presented 
measurements of $\sin2\beta_{\psi K}$ with unprecedented precision. The
results are $0.59\pm 0.14\pm 0.05$ (BaBar\cite{Aubert:2001nu}) and 
$0.99\pm 0.14\pm 0.06$ (Belle\cite{Abe:2001xe}), which when combined 
with earlier determinations lead to the new world average 
$\sin2\beta_{\psi K}=0.79\pm 0.10$. The expectation obtained from the 
global analysis of the unitarity triangle (leaving aside earlier 
$\sin2\beta_{\psi K}$ measurements) was $\sin2\beta=0.68\pm 0.21$ at 
95\% confidence level,\cite{Hocker:2001xe} in good agreement with the
new data.

The above discussion relies on the SM and could be upset if there 
existed a New Physics contribution to $B$--$\bar B$ mixing, or a new
contribution to the $b\to c\bar c s$ transition amplitude with a nonzero 
weak phase $\phi_{\rm NP}$.\cite{Fleischer:2001cw} The latter case is, 
however, rather unlikely, since such a nonstandard contribution could 
hardly compete with the large tree-level amplitude of the SM. Otherwise
there should be signals of direct CP violation in decays such as 
$B^\pm\to J/\psi\,K^\pm$, and there should be other $b\to q\bar q s$ 
New Physics contributions of similar strength ($\sim\lambda^2$), which 
would upset the phenomenology of charmless hadronic decays such as 
$B\to\pi K,\pi\pi$, etc. Hence, it appears safe to assume that even in 
the presence of New Physics the time-dependent CP asymmetry observed in 
$B\to J/\psi\,K$ decays measures the $B$--$\bar B$ mixing phase, so that 
$\sin2\beta_{\psi K}=-\sin\phi_d$. The good agreement of the measured 
phase with the SM prediction suggests that at least the dominant part of 
the $B$--$\bar B$ mixing phase is due to the phase of the CKM matrix 
element $V_{td}$. (This ignores the possibility of an accidental 
agreement made possible by a discrete ambiguity. In other words, there
could still be a large New Physics contribution to $B$--$\bar B$ mixing
such that $\phi_d\approx\pi+2\beta$.)

\section{The Quest for New Physics}

While we are amazed by the workings of the SM, some theorists will be 
disappointed by the fact that $\sin2\beta_{\psi K}$ does not show a 
hint for New Physics. In many extensions of the SM it would have been 
quite possible (sometimes even required) for the $B$--$\bar B$ mixing 
phase to differ from its SM 
value.\cite{Bergmann:2000ak,Kagan:2000wm,Silva:2001ym,Eyal:2000ys,Xing:2000cg} 
For instance, potentially large effects could arise in models with 
iso-singlet down-type quarks and tree-level flavor-changing neutral 
currents,\cite{Barenboim:2001zz} left-right symmetric models with 
spontaneous CP violation\cite{Ball:2000mb,Ball:2000yi,Bergmann:2001pm}
(which are now excluded by the data), and SUSY models with extended 
minimal flavor violation.\cite{Ali:2001ej} On the contrary, only small 
modifications of $\sin2\beta_{\psi K}$ are allowed in a class of models 
with so-called minimal flavor 
violation,\cite{Bergmann:2001pm,Buras:2001dm,Buras:2001xq} for which one 
can derive the bounds $0.52<\sin2\beta_{\psi K}<0.78$. 

Much like the stunning success of the SM in explaining electroweak
precision data, the observed lack of deviations from the predictions of 
the CKM model is, to some extent, surprising. Recall that the CKM 
mechanism neither explains the baryon asymmetry in the Universe nor
offers a clue as to why CP violation does not occur in the strong 
interactions. There are good arguments suggesting that the stability of 
the electroweak scale will be explained by some New Physics at or below 
the TeV scale. But virtually all extensions of the SM contain many new 
CP-violating couplings. For instance, a minimal, unconstrained (i.e., 
not fine-tuned) SUSY extension introduces 43 complex couplings in 
addition to the CKM phase. The fact that experiments have not shown 
any trace of nonstandard CP violation is puzzling and creates what one 
may call the ``CP problem''. It is not unlikely that the ``decoupling'' 
of nonstandard CP violation effects is linked to the decoupling of New 
Physics in the sector of electroweak symmetry breaking. In that sense, 
the $B$ factories offer a complementary strategy for probing TeV-scale 
physics. Like with the search for the explanation of electroweak 
symmetry breaking, the fact that we have not yet found New Physics in 
the flavor sector does not mean that it is not there, it just means we 
have to look harder. Hence, the strategy should be to keep searching for 
(probably small) deviations from the CKM paradigm with ever more precise 
measurements.

Given what we have learned about flavor physics in the kaon and beauty 
systems, there is still plenty of room for New Physics effects in both
mixing and weak decays, and there is reason to hope that departures from 
the predictions of the SM may be discovered soon. Following is a list of
options for discovering some potentially large New Physics effects:

\noindent
1. Check if the strength of $B_s$--$\bar B_s$ mixing is correctly 
predicted by the SM, i.e., confirm or disprove that 
$\Delta m_s\approx(17\pm 3)$\,ps$^{-1}$.

\noindent
2. Measure the CP-violating phase $\gamma=\mbox{arg}(V_{ub}^*)$ in the 
bottom sector and check if it agrees with the value inferred from the 
standard global analysis of the unitarity triangle using measurements of 
CP violation in the top sector. The current prejudice that $\gamma$ must 
be less than $90^\circ$ relies on the assumption that $B_s$--$\bar B_s$ 
mixing is not affected by New Physics. A first opportunity for probing 
$\gamma$ directly is offered by the analysis of charmless hadronic 
$B$ decays, as discussed below.

\noindent
3. Probe for New Physics in a variety of rare decay processes 
(proceeding through penguin and box diagrams in the SM). Some examples 
are:

\begin{itemize}
\item
Check if $\sin2\beta_{\phi K}=\sin2\beta_{\psi K}$. If not, this would
be clear evidence for New Physics in $b\to s\bar s s$ penguin 
transitions.\cite{Grossman:1997ke}

\item
Look for a direct CP asymmetry in $B\to X_s\gamma$ decays. Any signal 
exceeding the level of 1\% would be a clear sign of New Physics 
in radiative penguin decays.\cite{Kagan:1998bh}

\item
Check if $\gamma$ measured in the penguin-dominated decays $B\to\pi K$
agrees with $\gamma$ extracted from pure tree-processes such as 
$B\to D K$ and $B\to D^*\pi$.\cite{Grossman:1999av,Fleischer:2000jv}

\item
Look for New Physics in $B\to K l^+ l^-$ decays, for instance by testing
the prediction of a form-factor zero in the forward-backward 
asymmetry.\cite{Burdman:1998mk,Ali:2000mm,Beneke:2001at}
\end{itemize}

\noindent
4. Measure the branching ratios for the very rare kaon decays 
$K^\pm\to\pi^\pm\nu\bar\nu$ and $K_L\to\pi^0\nu\bar\nu$, which allow for
an independent construction of the unitarity 
triangle.\cite{Buchalla:1996fp}

\noindent
5. Search for New Physics in $D$--$\bar D$ mixing and charm weak decays.

\noindent
6. Continue to look for CP violation outside of flavor physics by 
probing electric dipole moments of the neutron and electron.

\section{Charmless Hadronic Decays}

After establishing the existence of a weak phase in the top sector by 
showing that $\mbox{Im}(V_{td}^2)\ne 0$, the next step in testing the 
CKM paradigm must be to explore the CP-violating phase in the bottom 
sector, $\gamma=\mbox{arg}(V_{ub}^*)$. In the SM the two phases are, 
of course, related to each other. However, as discussed above there is 
still much room for New Physics to affect the magnitude of flavor 
violations in both mixing and weak decays. 

\begin{figure}
\centerline{\includegraphics[width=0.48\textwidth]{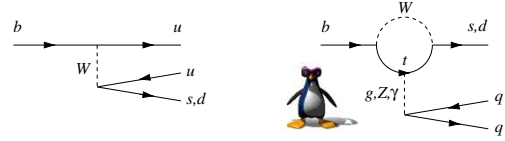}}
\caption{Tree and penguin topologies in charmless hadronic $B$ decays.}
\label{fig:treepeng}
\end{figure}

\begin{figure*}
\begin{center}
\includegraphics[width=0.8\textwidth]{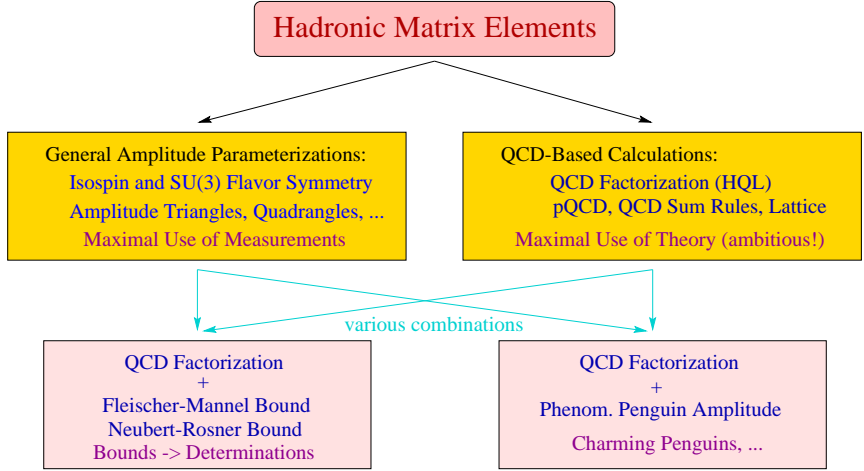}
\caption{Strategies used to determine hadronic matrix elements entering
charmless hadronic $B$ decays.}
\end{center}
\label{fig:strategies}
\end{figure*}

Common lore says that measurements of $\gamma$ are difficult. Several
theoretically clean determinations of this phase from pure tree
processes such as $B\to D\,K$\,\cite{Atwood:1997ci} and 
$B\to D^*\pi$\,\cite{Dunietz:1988bv} have been suggested, which are 
extremely challenging experimentally. Likewise, measurements of 
$\beta+\gamma=\pi-\alpha$ using isospin analysis in $B\to\pi\pi$ 
decays\cite{Gronau:1990ka} or Dalitz plot analysis in $B\to\pi\pi\pi$ 
decays\cite{Snyder:1993mx} are very difficult. It is more accessible 
experimentally to probe $\gamma$ via the sizeable tree--penguin 
interference in charmless hadronic decays such as $B\to\pi K$ and 
$B\to\pi\pi$. The basic decay topologies contributing to these modes 
are shown in Figure~\ref{fig:treepeng}. Experiment shows that the 
tree-to-penguin ratios in the two cases are roughly 
$|T/P|_{\pi K}\approx 0.2$ and $|P/T|_{\pi\pi}\approx 0.3$, indicating 
a sizeable amplitude interference. It is important that the relative 
weak phase between the two amplitudes can be probed not only via CP 
asymmetry measurements ($\sim\sin\gamma$), but also via measurements of 
CP-averaged branching fractions ($\sim\cos\gamma$). 

Extracting information about CKM parameters from the analysis of 
nonleptonic $B$ decays is a challenge to theory, since it requires some 
level of control over hadronic physics, including strong-interaction 
phases. Figure~\ref{fig:strategies} illustrates the two main strategies 
for tackling the problem of hadronic matrix elements: general amplitude 
parameterizations avoiding any dynamical input on one hand, and 
QCD-based calculations on the other. In the first approach, decay 
amplitudes are cataloged according to the flavor topologies that can 
contribute to a given decay process, such as tree topologies, penguin 
topologies, annihilation topologies, etc.\cite{Gronau:1994bn} The 
various topologies can be related to renormalization-group invariant 
combinations of operator matrix elements of the effective weak 
Hamiltonian,\cite{Buras:2000ra} but no attempt is made to calculate 
these matrix elements from first principles. Instead, isospin symmetry 
or, more generally, SU(3) flavor symmetry is used to obtain relations 
between the various amplitudes in different decay modes. Experimental 
data is then used to determine as many hadronic parameters as possible. 
This leads to the well-known amplitude triangle (and quadrangle) 
constructions, from which CP-violating phases can be extracted (modulo 
discrete ambiguities) in the limit of exact flavor 
symmetry.\cite{Fleischer:1997bv} QCD-based calculations of hadronic 
matrix elements are more ambitious in that they aim at an understanding 
of the underlying strong-interaction dynamics from first principles. 
Factorization theorems (such as the QCD factorization 
approach\cite{Beneke:1999br,Beneke:2000ry} and the hard-scattering 
approach\cite{Chang:1997dw,Keum:2001ph}) attack this problem by 
exploiting the heavy-quark limit. Other schemes, such as QCD sum rules 
and lattice QCD, are applicable to a wider class of processes, 
including hadronic decays of light mesons. Unfortunately, at present 
these approaches still face tremendous technical difficulties when 
attempting the calculation of nonleptonic decay amplitudes. A very 
promising strategy is to combine the results obtained using amplitude
parameterization with some dynamical information derived from QCD-based
calculations. For instance, in that way model-independent 
bounds\cite{Fleischer:1998um,Neubert:1998pt} on the CP-violating phase 
$\gamma$ can be turned into determinations of $\gamma$ that are subject 
to only very small theoretical uncertainties.

\section{QCD Factorization}

The statement that hadronic weak decay amplitudes simplify greatly in 
the heavy-quark limit $m_b\gg\Lambda_{\rm QCD}$ will not surprise those 
who have followed the dramatic advances in the theoretical understanding 
of $B$ physics during the past decade. Many areas of $B$ physics, from 
spectroscopy to exclusive semileptonic decays to inclusive rates and 
lifetimes, can now be systematically analyzed using heavy-quark 
expansions. Yet, the more complicated exclusive nonleptonic decays have
long resisted theoretical progress. The technical reason is that, 
whereas in most other applications of heavy-quark expansions one 
proceeds by integrating out heavy fields (leading to local operator 
product expansions), in the case of nonleptonic decays the large scale 
$m_b$ enters as the energy carried by light fields. Therefore, in 
addition to hard and soft subprocesses collinear degrees of freedom 
become important. This complicates the understanding of hadronic decay 
amplitudes using the language of effective field theory. (Yet, 
significant progress towards an effective field-theory description of 
nonleptonic decays has been made recently with the establishment of a 
``collinear--soft effective theory''.\cite{Bauer:2001yr} The reader is 
referred to these papers for more details on this development.) 

The importance of the heavy-quark limit is linked to the idea of color 
trans\-pa\-ren\-cy.\cite{Bjorken:1989kk,Dugan:1991de,Politzer:1991au} A 
fast-moving light meson (such as a pion) produced in a point-like source 
(a local operator in the effective weak Hamiltonian) decouples from soft 
QCD interactions. More precisely, the couplings of soft gluons to such a 
system can be analyzed using a multipole expansion, and the leading 
contribution (from the color dipole) is suppressed by a power of 
$\Lambda_{\rm QCD}/m_b$. The QCD factorization approach provides a 
systematic implementation of this idea.\cite{Beneke:1999br,Beneke:2000ry}
It yields rigorous results in the heavy-quark limit, which are valid to 
leading power in $\Lambda_{\rm QCD}/m_b$ but to all orders of 
perturbation theory. Having obtained control over nonleptonic decays in 
the heavy-quark limit is a tremendous advance. We are now able to talk 
about power corrections to a well-defined and calculable limiting case, 
which captures a substantial part of the physics in these complicated 
processes.

\begin{figure}[t]
\includegraphics[width=.46\textwidth]{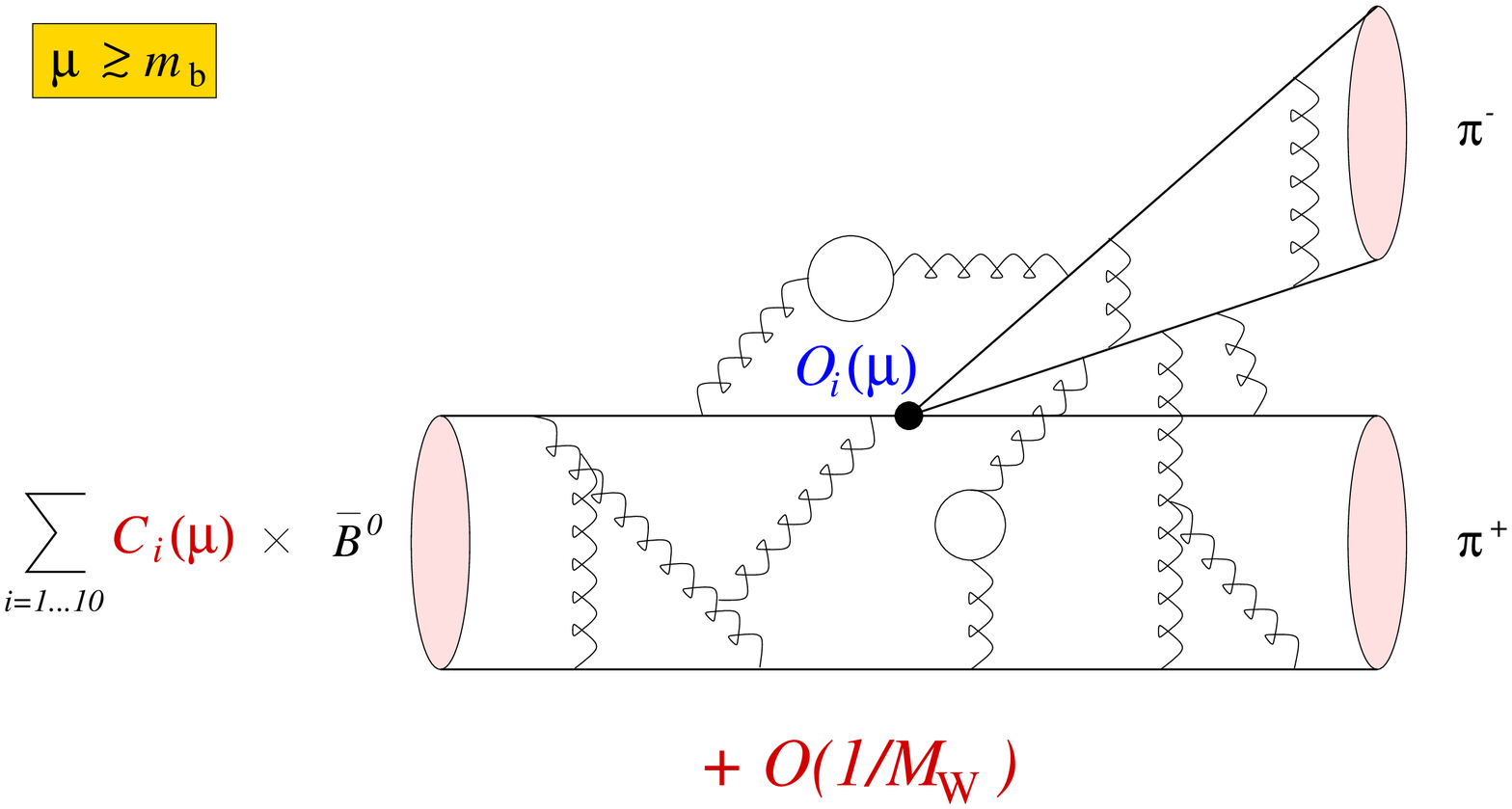}
\includegraphics[width=.48\textwidth]{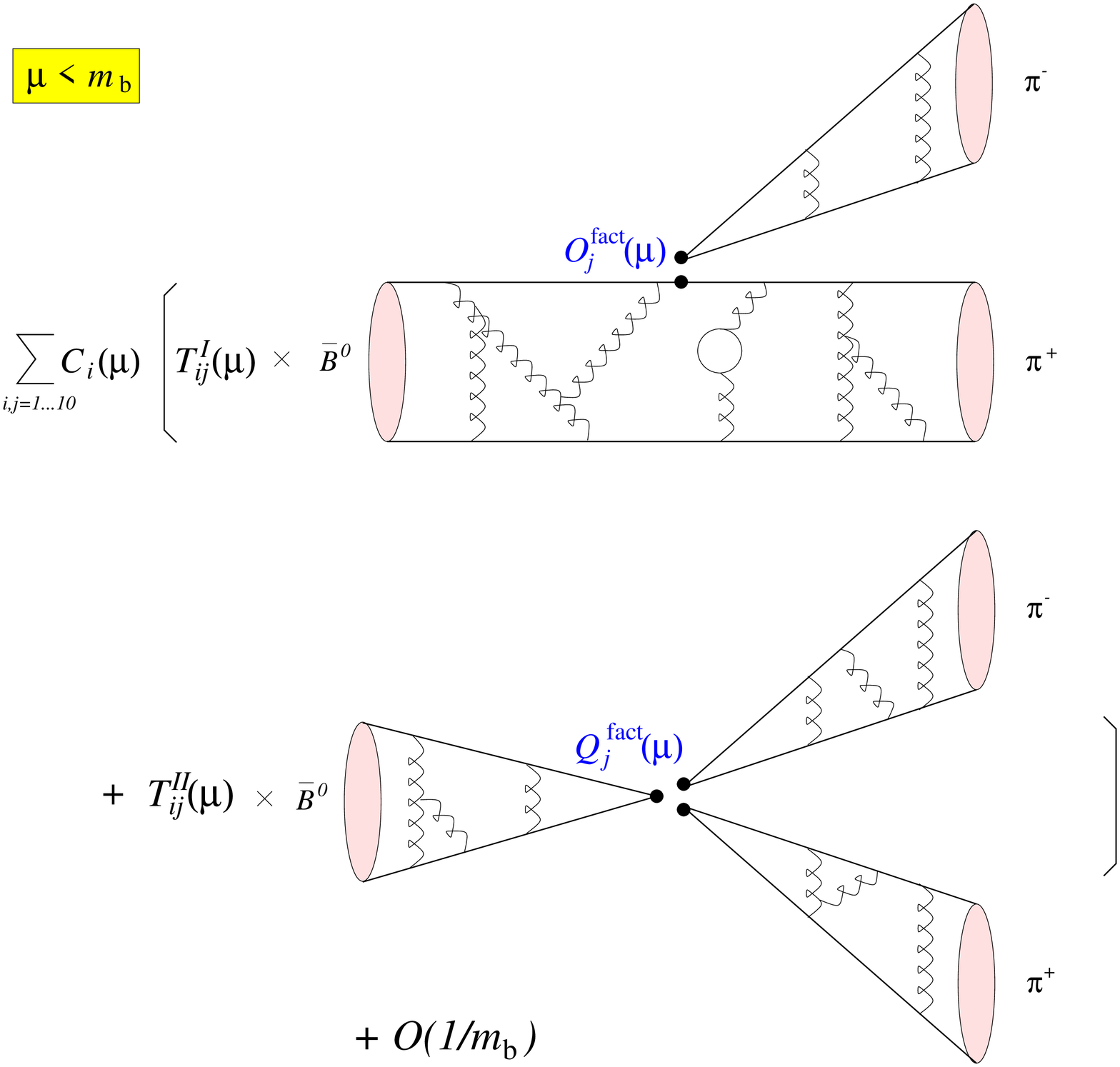}
\caption{Factorization of short- and long-distance contributions in 
hadronic $B$ decays. Upper: Factorization of short-distance effects into
Wilson coefficients of the effective weak Hamiltonian. Lower: 
Factorization of hard ``nonfactorizable'' gluon exchanges into 
hard-scattering kernels (QCD factorization).}
\label{fig:fact}
\end{figure}

The workings of QCD factorization can be illustrated with the cartoons 
shown in Figure~\ref{fig:fact}. The first graph shows the well-known 
concept of an effective weak Hamiltonian obtained by integrating out the 
heavy fields of the top quark and weak gauge bosons from the SM 
Lagrangian. This introduces new effective interactions mediated by local 
operators $O_i(\mu)$ (typically four-quark operators) multiplied by 
calculable running coupling constants $C_i(\mu)$ called Wilson 
coefficients. This reduction in complexity (nonlocal heavy particle 
exchanges $\to$ local effective interactions) is exact up to corrections 
suppressed by inverse powers of the heavy mass scales. The resulting 
picture at scales at or above $m_b$ is, however, still rather 
complicated, since gluon exchange is possible between any of the quarks 
in the external meson states. Additional simplifications occur when the 
renormalization scale $\mu$ is lowered below the scale $m_b$. Then color 
transparency comes to play and implies systematic cancellations of soft 
and collinear gluon exchanges. As a result, all ``nonfactorizable'' 
exchanges, i.e., gluons connecting the light meson at the ``upper'' 
vertex to the remaining mesons, are dominated by virtualities of order 
$m_b$ and can be calculated. Their effects are absorbed into a new set 
of running couplings $T_{ij}^{I,II}(\mu)$ called hard-scattering 
kernels, as shown in the two lower graphs. What remains are 
``factorized'' four-quark and six-quark operators $O_j^{\rm fact}(\mu)$ 
and $Q_j^{\rm fact}(\mu)$, whose matrix elements can be expressed in 
terms of form factors, decay constants and light-cone distribution 
amplitudes. As before, the reduction in complexity (local four-quark 
operators $\to$ ``factorized'' operators) is exact up to corrections 
suppressed by inverse powers of the heavy scale, now set by the 
$b$-quark mass.

The factorization formula is valid in all cases where the meson at the
``upper'' vertex is light, meaning that its mass is much smaller than 
the $b$-quark mass. The second term in the factorization formula (the 
term involving ``factorized'' six-quark operators) gives a 
power-suppressed contribution when the final-state meson at the 
``lower'' vertex is a heavy meson (i.e., a charm meson), but its 
contribution is of leading power if this meson is also light. 

Factorization is a property of decay amplitudes in the heavy-quark 
limit. Given the magnitudes of ``nonfactorizable'' effects seen in kaon, 
charm and beauty decays, there can be little doubt about the relevance 
of the heavy-quark limit to understanding nonleptonic 
processes.\cite{Neubert:2001sj} Yet, for phenomenological applications 
it is important to explore the structure of at least the leading 
power-suppressed corrections. While no complete classification of such 
corrections has been achieved to date, several classes of 
power-suppressed terms have been analyzed and their effects estimated. 
They include ``chirally-enhanced'' power corrections,\cite{Beneke:1999br}
weak annihilation 
contributions,\cite{Beneke:2000ry,Cheng:2001hv,Beneke:2001ev} and power 
corrections due to nonfactorizable soft gluon 
exchange.\cite{Khodjamirian:2001mi,Burrell:2001pf,Becher:2001hu} With 
the exception of the ``chirally-enhanced'' terms, no unusually large 
power corrections (i.e., corrections exceeding the naive expectation of 
5--10\%) have been identified so far. Nevertheless, it is important to 
refine and extend the estimates of power corrections. Fortunately, the 
QCD factorization approach makes many testable predictions. Ultimately, 
therefore, the data will give us conclusive evidence on the relevance of
power-suppressed effects. Many tests can already be performed using 
existing data.

\subsection{Tests of Factorization in $B\to D^{(*)} L$} 

In $B$ decays into a heavy--light final state, when the light meson is 
produced at the ``upper'' vertex, the factorization formula assumes its
simplest form. Then only the form factor term (the first graph in the 
lower portion of Figure~\ref{fig:fact}) contributes at leading power. 
This is also the place where QCD factorization is best established 
theoretically. The systematic cancellation of soft and collinear 
singularities has been demonstrated to all orders in perturbation 
theory,\cite{Beneke:2000ry,Bauer:2001cu} and in the ``large-$\beta_0$ 
limit'' of QCD it has been shown that the hard-scattering kernels are 
free of power-like endpoint singularities as one of the quarks in the 
light meson becomes a soft parton.\cite{Becher:2001hu} (It is still an 
open question whether such a smooth behavior persists in higher orders 
of full QCD.)

In the case of the decays $\bar B^0\to D^{(*)+} L^-$, where $L$ denotes 
a light meson, the flavor content of the final state is such that the 
light meson can only be produced at the ``upper'' vertex, so 
factorization applies. One finds that process-dependent 
``nonfactorizable'' corrections from hard gluon exchange, though 
present, are numerically very small. All nontrivial QCD effects in the 
decay amplitudes are then described by a quasi-universal coefficient 
$|a_1(D^{(*)}L)|=1.05\pm 0.02+O(\Lambda_{\rm QCD}/
m_b)$.\cite{Beneke:2000ry} For a given decay channel this coefficient 
can be determined experimentally from the ratio\cite{Bjorken:1989kk}
\[
   \frac{\Gamma(\bar B^0\to D^{*+}L^-)}
        {d\Gamma(\bar B^0\to D^{*+}l^-\nu)/dq^2 \big|_{q^2=m_L^2}}
\]
\vspace{-0.5truecm}
\[
   = 6\pi^2 |V_{ud}|^2 f_L^2\,|a_1(D^{(*)}L)|^2 \,.
\]
Using CLEO data one obtains $|a_1(D^*\pi)|=1.08\pm 0.07$, 
$|a_1(D^*\rho)|=1.09\pm 0.10$, and $|a_1(D^* a_1)|=1.08\pm 0.11$, in 
good agreement with the theoretical prediction. This is a first 
indication that power corrections in these modes are under control, but 
more precise data are required for a firm conclusion. For other tests 
of factorization in $B$ decays to heavy--light final states the reader 
is referred to recent 
literature.\cite{Beneke:2000ry,Ligeti:2001dk,Diehl:2001xe}

The experimental observation of unexpectedly large rates for 
color-suppressed decays\cite{CLEO,Abe:2001pd} such as 
$\bar B^0\to D^{0(*)}\pi^0$ has attracted some attention. QCD 
factorization does not allow us to calculate the amplitudes for these 
processes in a reliable way. It predicts that these amplitudes are 
power-suppressed with respect to the corresponding 
$\bar B^0\to D^{+(*)}\pi^-$ amplitudes, but only by one power of 
$\Lambda_{\rm QCD}/m_c$. Specifically, the prediction is that a certain 
ratio of isospin amplitudes approaches unity in the heavy-quark limit: 
$A_{1/2}/(\sqrt2 A_{3/2})=1+O(\Lambda_{\rm QCD}/m_c)$. This scaling law 
is respected by the experimental data, which give 
$A_{1/2}/(\sqrt2 A_{3/2})=(0.70\pm 0.11)\,e^{\pm i(27\pm 7)^\circ}$ for
$B\to D\,\pi$ and $(0.72\pm 0.08)\,e^{\pm i(21\pm 8)^\circ}$ for
$B\to D^*\pi$.\cite{Neubert:2001sj} A rough theoretical estimate of the 
amplitude ratio, $A_{1/2}/(\sqrt2 A_{3/2})\approx 
0.75\,e^{-15^\circ\,i}$, had been obtained prior to the observation of 
the color-suppressed decays.\cite{Beneke:2000ry} It anticipated the 
correct order of magnitude of the deviation from the heavy-quark limit. 

\subsection{Tests of Factorization in $B\to K^*\gamma$} 

The QCD factorization approach not only applies to nonleptonic decays, 
but also to other exclusive processes such as $B\to V\gamma$ and 
$B\to V\,l^+ l^-$, where $V=K^*,\rho,\dots$ is a vector 
meson.\cite{Beneke:2001at,Bosch:2001gv} The resulting factorization 
formula is similar (but simpler) to that for $B$ decays into two light 
mesons. The study of exclusive radiative transitions therefore not 
only extends the range of applicability of the method, it also provides 
a new testing ground for the factorization idea.

Interestingly, the analysis of isospin-breaking effects in radiative $B$ 
decays, as measured by the 
quantity\cite{Coan:2000kh,Ushiroda:2001sb,BaBar}
\begin{eqnarray}
   \Delta_{0-} &\equiv&
    \frac{\Gamma(\bar B^0\to\bar K^{*0}\gamma)
          -\Gamma(B^-\to\bar K^{*-}\gamma)}
         {\Gamma(\bar B^0\to\bar K^{*0}\gamma)
          +\Gamma(B^-\to\bar K^{*-}\gamma)} \nonumber\\
   &=& 0.11\pm 0.07 \,, \nonumber
\end{eqnarray}
gives a direct probe of power corrections to the factorization formula,
since such effects are absent in the heavy-quark limit. A theoretical 
analysis of the leading power-suppressed contributions gives 
$\Delta_{0-}=(8.0_{\,-\,3.2}^{\,+\,2.1})\%\times
(0.3/T_1^{B\to K^*})$,\cite{Kagan:2001zk} where $T_1^{B\to K^*}\sim 0.3$ 
is a $B\to K^*$ form factor. The largest contribution comes from an 
annihilation contribution involving the penguin operator $O_6$ in the 
effective weak Hamiltonian. As a result, the quantity $\Delta_{0-}$ is a 
sensitive probe of the magnitude and sign of the ratio $C_6/C_{7\gamma}$ 
of Wilson coefficients.

The theoretical prediction for $\Delta_{0-}$ is in agreement with the 
current experimental value. If this agreement persists as the data 
become more precise, this would not only test the penguin sector of the 
effective weak Hamiltonian but also provide a quantitative test of 
factorization at the level of power corrections.

\subsection{$\!\!$Tests of Factorization in $B\!\to\!\pi K,\pi\pi$} 

The factorization formula for $B$ decays into two light mesons is more
complicated because of the presence of the two types of contributions
shown in the lower graph in Figure~\ref{fig:fact}. The finding that 
these two topologies contribute at the same power in 
$\Lambda_{\rm QCD}/m_b$ is nontrivial\cite{Beneke:2001ev} and relies 
on the heavy-quark scaling law $F^{B\to L}(0)\sim m_b^{-3/2}$ for 
heavy-to-light form 
factors,\cite{Chernyak:1990ag,Ali:1994vd,Bagan:1998bp} which is 
established less rigorously than the corresponding scaling law for 
heavy-to-heavy form factors. In the QCD factorization approach the 
kernels $T_{ij}^I(\mu)$ are of order unity, whereas the kernels 
$T_{ij}^{II}(\mu)$ contribute first at order $\alpha_s$. Numerically, 
the latter ones give corrections of about 10--20\% with respect to the 
leading terms. Therefore, the scaling laws that form the basis of the 
QCD factorization formula appear to work well empirically.

The factorization formula for $B$ decays into two light mesons can be 
tested best by using decays that have negligible amplitude interference. 
In that way any sensitivity to the value of the weak phase $\gamma$ is 
avoided. For a complete theoretical control over charmless hadronic 
decays one must control the magnitude of the tree topologies, the 
magnitude of the penguin topologies, and the relative strong-interaction 
phases between trees and penguins. It is important that these three key 
features can be tested separately. Once these tests are conclusive (and 
assuming they are successful), factorization can be used to constrain 
the parameters of the unitarity triangle.

\subsection*{Magnitude of the Tree Amplitude}
The magnitude of the leading $B\to\pi\pi$ tree amplitude can be probed in
the decays $B^\pm\to\pi^\pm\pi^0$, which to an excellent approximation 
do not receive any penguin contributions. The QCD factorization approach 
makes an absolute prediction for the corresponding branching 
ratio,\cite{Beneke:2001ev}
\[
   \mbox{Br}(B^\pm\to\pi^\pm\pi^0)
   = \left[ \frac{|V_{ub}|}{0.0035}\,\frac{F_0^{B\to\pi}(0)}{0.28}
   \right]^2
\]
\vspace{-0.5truecm}
\[
   \times \Big[ 5.3_{\,-0.4}^{\,+0.8}\,(\mbox{pars.})
   \pm 0.3\,(\mbox{power}) \Big] \cdot 10^{-6} \,,
\]
which compares well with the experimental result
$(5.7\pm 1.5)\times 10^{-6}$ (see the table in Figure~\ref{fig:UTfit}
for a compilation of the experimental data on charmless hadronic 
decays). The theoretical uncertainties quoted are due to input parameter 
variations and the modeling of power corrections. An additional large 
uncertainty comes from the present error on $|V_{ub}|$ and the 
semileptonic $B\to\pi$ form factor. The sensitivity to these quantities 
can be eliminated by taking the ratio 
\[
   \frac{\Gamma(B^\pm\to\pi^\pm\pi^0)}
        {d\Gamma(\bar B^0\to\pi^+ l^-\bar\nu)/dq^2|_{q^2=0}}
\]
\vspace{-0.5truecm}
\[
   = 3\pi^2 f_\pi^2 \hspace{-0.9cm}
   \underbrace{|a_1^{(\pi\pi)}+a_2^{(\pi\pi)}|^2
               }_{1.33_{\,-0.11}^{\,+0.20}\,(\mbox{pars.})\pm 0.07\,
               (\mbox{power})} \hspace{-0.9cm} 
   = (0.68_{\,-0.06}^{\,+0.11})\,\mbox{GeV}^2 .
\]
This prediction includes a sizeable ($\sim 25\%$) contribution of the 
hard-scattering term in the factorization formula (the last graph in 
Figure~\ref{fig:fact}). Unfortunately, this ratio has not yet been 
measured experimentally.

\subsection*{Magnitude of the $T/P$ Ratio}
The magnitude of the leading $B\to\pi K$ penguin amplitude can be probed 
in the decays $B^\pm\to\pi^\pm K^0$, which to an excellent approximation 
do not receive any tree contributions. Combining it with the measurement
of the tree amplitude just described, a tree-to-penguin ratio can be 
determined via the relation
\[
   \varepsilon_{\rm exp}\! = \!\left| \frac{T}{P} \right|\!
   = \mbox{tan}\theta_C\,\frac{f_K}{f_\pi}\!\left[ 
   \frac{2\mbox{Br}(B^\pm\to\pi^\pm\pi^0)}
        {\mbox{Br}(B^\pm\to\pi^\pm K^0)} \right]^{\frac12} \!\!.
\]
The present experimental value $\varepsilon_{\rm exp}=0.223\pm 0.034$ is 
in good agreement with the theoretical prediction 
$\varepsilon_{\rm th}=0.24\pm 0.04\,(\mbox{pars.})\pm 0.04\,
(\mbox{power})\pm 0.05\,(V_{ub})$,\cite{Beneke:2001ev} which is 
independent of form factors but proportional to $|V_{ub}/V_{cb}|$. This 
is a highly nontrivial test of the QCD factorization approach. Recall 
that, when the first measurements of charmless hadronic decays appeared, 
several authors remarked that the penguin amplitudes were much larger 
than expected based on naive factorization models. We now see that QCD 
factorization reproduces naturally (i.e., for central values of all 
input parameters) the correct magnitude of the tree-to-penguin ratio. 
This observation also shows that there is no need to supplement the QCD 
factorization predictions in an ad hoc way by adding enhanced 
phenomenological penguin amplitudes, such as the ``nonperturbative 
charming penguins'' introduced some time ago.\cite{Ciuchini:1997hb} 
(The effects of charming penguins can be parameterized in terms of a 
``bag parameter'' $\hat B_1=(0.13\pm 0.02)\,e^{i(188\pm 82)^\circ}$ 
fitted to the data on charmless decays.\cite{Ciuchini:2001pk} By 
definition, this parameter contains the contribution from the 
perturbative charm loop, which is calculable in QCD factorization. Using 
the factorization approach one finds that $\hat B_1^{\rm fact}
=(0.09_{\,-0.02-0.02}^{\,+0.03+0.04})\,e^{i(185\pm 3\pm 21)^\circ}$, 
where the errors are due to input parameter variations and the estimate 
of power corrections. The perturbative contribution to the central value 
is 0.08; the remaining 0.01 is mainly due to weak annihilation. Hence, 
within errors QCD factorization can account for the ``charming penguin 
bag parameter'', which is in fact dominated by short-distance physics.)

\begin{table*}[t]
\vspace{0.3truecm}
\begin{center}
\caption{Direct CP asymmetries in $B\to\pi K$ decays\label{tab:ACPs}}
\begin{tabular}{c|c|ccc}
\hline
 & Experiment & \multicolumn{3}{c}{Theory} \\
 & \cite{Chen:2000hv,Abe:2001hs,Aubert:2001hs,Aubert:2001qj}
 & Beneke et al.\cite{Beneke:2001ev}
 & Keum et al.\cite{Keum:2001ph}
 & Ciuchini et al.\cite{Ciuchini:1997hb}
 \\
\hline
$A_{\rm CP}(\pi^+ K^-)$~(\%) & $-4.8\pm 6.8$ & $5\pm 9$ & $-18$
 & $\pm(17\pm 6)$ \\
$A_{\rm CP}(\pi^0 K^-)$~(\%) & $-9.6\pm 11.9$ & $7\pm 9$ & $-15$
 & $\pm(18\pm 6)$ \\
$A_{\rm CP}(\pi^-\bar K^0)$~(\%) & $-4.7\pm 13.9$ & $1\pm 1$ & $-2$
 & $\pm(3\pm 3)$ \\
\hline
\end{tabular}
\end{center}
\end{table*}

\subsection*{Strong Phase of the $T/P$ Ratio}
The QCD factorization approach predicts that strong-interaction phases 
in most charmless hadronic $B$ decays are parametrically suppressed in 
the heavy-quark limit, i.e., 
$\phi_{\rm st}=O[\alpha_s(m_b),\Lambda_{\rm QCD}/m_b]$. This implies
small direct CP asymmetries since, e.g., $A_{\rm CP}(\pi^+ K^-)\simeq 
-2\,|\frac{T}{P}|\sin\gamma\,\sin\phi_{\rm st}$. The suppression results 
as a consequence of systematic cancellations of soft contributions, 
which are missed in phenomenological models of final-state interactions. 
In many other schemes the strong-interaction phases are predicted to be 
much larger, and therefore larger CP asymmetries are expected. 
Table~\ref{tab:ACPs} shows that first experimental data provide no 
evidence for large direct CP asymmetries in $B\to\pi K$ decays. However, 
the errors are still too large to draw a definitive conclusion that 
would allow us to distinguish between different theoretical predictions.

\subsection{Remarks on Sudakov Logarithms}

In recent years, Li and collaborators have proposed an alternative
scheme for calculating nonleptonic $B$ decay amplitudes based on a 
perturbative hard-scattering approach.\cite{Chang:1997dw,Keum:2001ph} 
From a conceptual point of view, the main difference between QCD 
factorization and this so-called pQCD approach lies in the latter's 
assumption that Sudakov form factors effectively suppress soft-gluon 
exchange in diagrams such as those shown in Figure~\ref{fig:fact}. As a 
result, the $B\to\pi$ and $B\to K$ form factors are assumed to be 
perturbatively calculable. This changes the counting of powers of 
$\alpha_s$. In particular, the nonfactorizable gluon-exchange diagrams 
included in the QCD factorization approach, which are crucial in order 
to cancel the scale and scheme-dependence in the predictions for the 
decay amplitudes, are formally of order $\alpha_s^2$ in the pQCD scheme 
and consequently are left out. Thus, to the considered order there are 
no loop graphs that could give rise to strong-interaction phases in that 
scheme. (However, large phases are claimed to arise from on-shell poles 
of massless propagators in tree diagrams.\cite{Keum:2001ph} Because 
these phases are dominated by soft physics, the prediction of large 
direct CP asymmetries in the pQCD approach rests on assumptions that are 
strongly model dependent.)

The assumption of Sudakov suppression in hadronic $B$ decays is 
questionable, because the relevant scale 
$Q^2\sim m_b\Lambda_{\rm QCD}\sim 1$\,GeV$^2$ is not that large for 
realistic $b$-quark masses. Indeed, one finds that the pQCD calculations 
are very sensitive to details of the $p_\perp$ dependence of the wave 
functions.\cite{Descotes-Genon:2001hm} This sensitivity to hadronic 
physics invalidates the original assumption of an effective suppression 
of soft contributions. (The argument just presented leaves open the 
conceptual question whether Sudakov logarithms are relevant in the 
asymptotic limit $m_b\to\infty$. This question has not yet been 
answered in a satisfactory way.)

\boldmath
\section{Constraints in the $(\bar\rho,\bar\eta)$ Plane}
\unboldmath

The QCD factorization approach, combined with a conservative estimate of 
power corrections, offers several new strategies to derive constraints 
on CKM parameters.\cite{Beneke:2001ev} Some of these strategies will be 
illustrated below. Note that the applications of QCD factorization are 
not limited to computing branching ratios. The approach is also useful 
in combination with other ideas based on flavor symmetries and amplitude 
relations. In this way, strategies can be found for which the residual 
hadronic uncertainties are simultaneously suppressed by three small 
parameters, since they vanish in the heavy-quark limit 
($\sim\Lambda_{\rm QCD}/m_b$), the limit of SU(3) flavor symmetry 
($\sim (m_s-m_q)/\Lambda_{\rm QCD}$), and the large-$N_c$ limit 
($\sim1/N_c$).

\subsection{Extraction of $\,\gamma$ with Minimal Theory Input}

\begin{figure}[t]
\centerline{\includegraphics[width=0.48\textwidth]{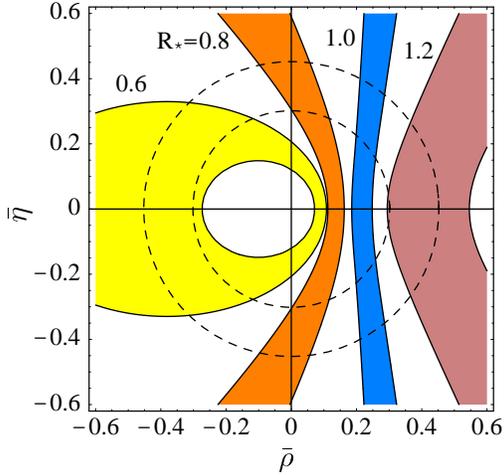}}
\caption{Allowed regions in the $(\bar\rho,\bar\eta)$ plane 
corresponding to $\varepsilon_{\rm exp}=0.22$ and different values of 
the ratio $R_*$. The widths of the bands reflect the theoretical 
uncertainty. The current experimental values are 
$\varepsilon_{\rm exp}=0.22\pm 0.03$ and $R_*=0.71\pm 0.14$.}
\label{fig:NR}
\end{figure}

Some years ago, Rosner and the present author have derived a bound on 
$\gamma$ by combining measurements of the ratios 
$\varepsilon_{\rm exp}=|T/P|$ and 
$R_*=\frac12\,\Gamma(B^\pm\to\pi^\pm K^0)/\Gamma(B^\pm\to\pi^0 K^\pm)$ 
with the fact that for an arbitrary strong-interaction phase
$-1\le\cos\phi_{\rm st}\le 1$.\cite{Neubert:1998pt} The 
model-independent observation that $\cos\phi_{\rm st}=1$ up to 
second-order corrections to the heavy-quark limit can be used to turn 
this bound into a determination of $\gamma$ (once $|V_{ub}|$ is known). 
The resulting constraints in the $(\bar\rho,\bar\eta)$ plane, obtained
under the conservative assumption that $\cos\phi_{\rm st}>0.8$ 
(corresponding to $|\phi_{\rm st}|<37^\circ$), are shown in 
Figure~\ref{fig:NR} for several illustrative values of the ratio $R_*$. 
Note that for $0.8<R_*<1.1$ (the range preferred by the SM) the 
theoretical uncertainty reflected by the widths of the bands is smaller 
than for any other constraint on $(\bar\rho,\bar\eta)$ except for the 
one derived from the $\sin2\beta_{\psi K}$ measurement. With present 
data the SM is still in good shape, but it will be interesting to see 
what happens when the experimental errors are reduced.

\begin{figure}
\centerline{\includegraphics[width=0.48\textwidth]{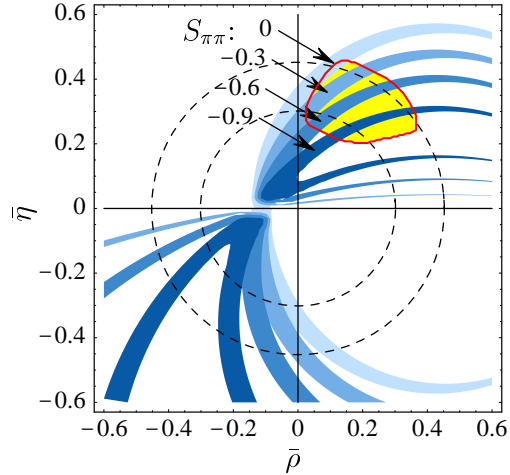}}
\caption{Allowed regions in the $(\bar\rho,\bar\eta)$ plane 
corresponding to different values of $S_{\pi\pi}$. The widths of the 
bands reflect the theoretical uncertainty. The corresponding bands for 
positive $S_{\pi\pi}$ are obtained by a reflection about the $\bar\rho$ 
axis. The bounded light area shows the allowed region obtained from the 
standard analysis of the unitarity triangle.\protect\cite{Hocker:2001xe}}
\label{fig:sin2a}
\end{figure}

\subsection{Determination of sin\,2$\alpha$}

With the help of QCD factorization it is possible to control the 
``penguin pollution'' in the time-dependent CP asymmetry in 
$B\to\pi^+\pi^-$ decays, defined such that 
$S_{\pi\pi}=\sin2\alpha\cdot[1+O(P/T)]$. This is illustrated in 
Figure~\ref{fig:sin2a}, which shows the constraints imposed by a 
measurement of $S_{\pi\pi}$ in the $(\bar\rho,\bar\eta)$ plane. Even a 
result for $S_{\pi\pi}$ with large experimental errors would imply a 
useful constraint on the unitarity triangle. A first, preliminary 
measurement of the asymmetry has been presented by the BaBar 
Collaboration at this conference.\cite{Aubert:2001qj} Their result is 
$S_{\pi\pi}=0.03_{\,-0.56}^{\,+0.53}\pm 0.11$.

\subsection{Global Fit to the $B\to\pi K,\pi\pi$ Branching Ratios}

\begin{figure}
\centerline{\includegraphics[width=0.48\textwidth]{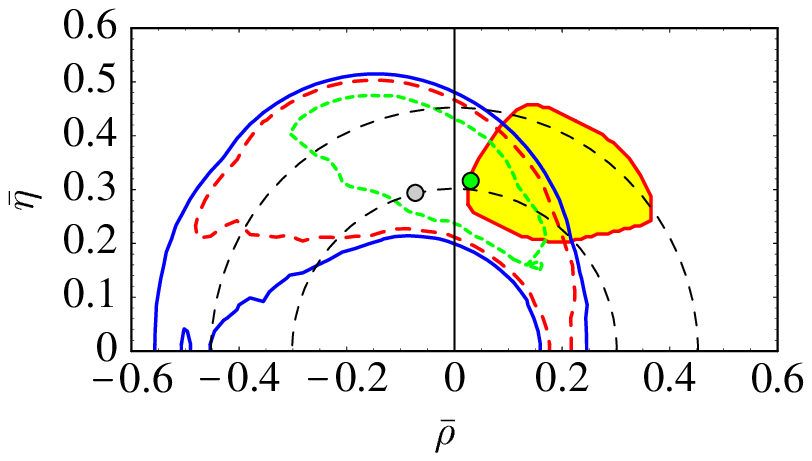}}
\centerline{
\begin{tabular}{l|c|c}
\hline
Decay Mode & Best Fit & Exp.\ Average \\
\hline
$B^0\to\pi^+\pi^-$ & 4.6 & $4.1\pm 0.9$ \\
$B^\pm\to\pi^\pm\pi^0$ & 5.3 & $5.7\pm 1.5$ \\
\hline
$B^0\to\pi^\mp K^\pm$ & 17.9 & $17.3\pm 1.5$ \\
$B^\pm\to\pi^0 K^\pm$ & 11.3 & $12.0\pm 1.6$ \\
$B^\pm\to\pi^\pm K^0$ & 17.7 & $17.2\pm 2.6$ \\
$B^0\to\pi^0 K^0$ & 7.1 & $10.4\pm 2.6$ \\
\hline
\end{tabular}}
\vspace{0.3truecm}
\caption{95\% (solid), 90\% (dashed) and 68\% (short-dashed) confidence 
level contours in the $(\bar\rho,\bar\eta)$ plane obtained from a global 
fit of QCD factorization results to the CP-averaged $B\to\pi K,\pi\pi$ 
branching fractions. The dark dot shows the overall best fit; the light 
dot indicates the best fit for the default parameter set. The table 
compares the best fit values for the branching fractions (in units of 
$10^{-6}$) with the world average 
data.\protect\cite{Aubert:2001hs,Cronin-Hennessy:2000kg,Abe:2001nq}}
\label{fig:UTfit}
\end{figure}

Various ratios of CP-averaged $B\to\pi K,\pi\pi$ branching fractions 
exhibit a strong dependence on $\gamma$ and $|V_{ub}|$, or equivalently,
on the parameters $\bar\rho$ and $\bar\eta$ of the unitarity triangle. 
From a global analysis of the experimental data it is possible to derive 
constraints in the $(\bar\rho,\bar\eta)$ plane in the form of regions 
allowed at various confidence levels. The results of such an analysis 
are shown in Figure~\ref{fig:UTfit}. The best fit of the QCD 
factorization theory to the data yields an excellent $\chi^2/n_{\rm dof}$ 
of about 0.5. (We disagree with the implementation of our approach 
presented in recent work by Ciuchini et al.\cite{Ciuchini:2001pk} and, 
in particular, with the numerical results labeled ``BBNS'' in Table~II 
of that paper, which led the authors to the conclusion that the ``theory 
of QCD factorization ... is insufficient to fit the data''. Even 
restricting $(\bar\rho,\bar\eta)$ to lie within the narrow ranges
adopted by these authors, one finds parameter sets for which QCD 
factorization fits the data with a $\chi^2/n_{\rm dof}$ of less than 
1.5.) 

The results of this global fit are compatible with the standard CKM fit 
using semileptonic decays, $K$--$\bar K$ mixing, and $B$--$\bar B$ mixing
($|V_{ub}|$, $|V_{cb}|$, $\epsilon_K$, $\Delta m_d$, $\Delta m_s$, 
$\sin 2\beta_{\psi K}$), although the fit prefers a slightly larger 
value of $\gamma$ and a smaller value of $|V_{ub}|$. The combination 
of the results from rare hadronic $B$ decays with $|V_{ub}|$ from 
semileptonic decays excludes $\bar\eta=0$ at 95\% confidence level, thus 
showing first evidence for the existence of a CP-violating phase in the 
bottom sector. Very soon, when the data become more precise, this will 
provide a powerful test of the CKM paradigm.

\section{Outlook}

The field of $B$ physics and CP violation is more lively and fascinating
than ever. This year's discovery of CP violation in the $B$ system, 
combined with the recent discovery of direct CP violation in $K$ decays,
are outstanding achievements and a triumph for the SM. They establish 
the CKM mechanism as the dominant source of CP violation in hadronic 
weak decays.

Yet, searches for deviations form the CKM paradigm are well motivated 
and must be continued with ever higher level of precision. Some key 
measurements that can be performed in the near future include the
observation of $B_s$--$\bar B_s$ mixing, the measurement of the
CP-violating phase $\gamma$ in the bottom sector, and the discovery of 
the time-dependent CP asymmetry in $B\to\pi^+\pi^-$ decays. On the 
longer term, the main focus of $B$ physics should be on a systematic, 
detailed study of rare decay processes.

The QCD factorization approach provides the theoretical framework for a 
systematic analysis of hadronic and radiative exclusive $B$ decay 
amplitudes based on the heavy-quark expansion. This theory has already 
passed successfully several nontrivial tests, and will be tested more 
thoroughly with more precise data. Ultimately, this may lead to 
theoretical control over a vast variety of exclusive $B$ decays, giving 
us new constraints on the unitarity triangle.

If the CKM mechanism remains to stand ever more precise experimental 
test we will be facing a new decoupling problem, whose resolution may be 
linked to whatever New Physics there is to discover at the TeV scale.

\section*{Acknowledgments}
I wish to thank the organizers of {\em LP01\/} for the invitation to 
deliver this talk and for their support. I am grateful to T.~Becher, 
M.~Beneke, G.~Buchalla, Y.~Grossman, A.~Kagan, B.~Pecjak, A.~Petrov,
J.~Rosner, and C.~Sachrajda for many enjoyable collaborations on work 
relevant to this talk. This work was supported in part by the National 
Science Foundation.


\end{document}